\documentclass[preprint,showpacs,showkeys]{revtex4}

\usepackage{amsmath,bm} 
\usepackage{amsfonts} 
\usepackage{graphicx} 
\usepackage{mathrsfs} 
\usepackage{bbm}
\usepackage[cal=boondox,scr=boondoxo]{mathalfa}
\usepackage[bbgreekl]{mathbbol}
\usepackage{hyperref}

\DeclareSymbolFontAlphabet{\mathbb}{AMSb}
\DeclareSymbolFontAlphabet{\mathbbl}{bbold}

\begin{document}
\preprint{\textsl{KPOP}$\mathscr{E}$-2020-04}

\title{Bra-Ket Representation of the Inertia Tensor}

\author{U-Rae Kim}
\affiliation{\textsl{KPOP}$\mathscr{E}$ Collaboration, Department of Physics, Korea University, Seoul 02841, Korea}

\author{Dohyun Kim}
\affiliation{\textsl{KPOP}$\mathscr{E}$ Collaboration, Department of Physics, Korea University, Seoul 02841, Korea}

\author{Jungil Lee}
\email{jungil@korea.ac.kr} 
\thanks{Director of the Korea Pragmatist Organization for Physics Education (\textsl{KPOP}$\mathscr{E}$)}
\affiliation{\textsl{KPOP}$\mathscr{E}$ Collaboration, Department of Physics, Korea University, Seoul 02841, Korea}

\date{\today}

\begin{abstract}
We employ Dirac's bra-ket notation to define the
inertia tensor operator that is independent of the choice of bases or coordinate system. 
The principal axes and the corresponding principal values for the elliptic plate
are determined only based on the geometry. 
By making use of a general symmetric tensor operator, 
we develop a method of diagonalization that is convenient and intuitive in determining the eigenvector.
We demonstrate that the bra-ket approach greatly simplifies the computation of the inertia tensor 
with an example of an $N$-dimensional ellipsoid.
The exploitation of the bra-ket notation to compute the inertia tensor in classical mechanics 
should provide undergraduate students with 
a strong background necessary to deal with abstract quantum mechanical problems.
\end{abstract}


\maketitle 

\section{Introduction}
The inertia tensor is one of the essential ingredients in classical mechanics
with which one can investigate the rotational properties of rigid-body
motion \cite{CM}. The symmetric nature of the rank-2 Cartesian tensor 
guarantees that it is described by three fundamental parameters
called the principal moments of inertia $I_i$, each of which is the moment of inertia
along a principal axis. The principal axes are orthogonal so that
they construct a natural choice of the Cartesian coordinate system. Additional
three parameters of the symmetric tensor can be allocated to the
rotation from the natural coordinate system of the principal axes to
an arbitrary coordinate system. Thus, the tensor is one of the best 
tools with which students can train on how to deal with the eigenvalue
problem of a Hermitian matrix. The underlying mathematical structure
of this problem is closely related to the normal modes. It can
further be extended to the Hilbert space in the continuum limit~\cite{QM}.
Hence, we believe that teaching inertia tensors is of great importance
in giving students a strong background for their
future study of quantum mechanics.

Many pedagogical materials involving the general and the inertia tensors are available \cite{Dixon-1951,Rogers-2004,Rogers-2005,Sivardiere-2005,Battaglia-2013,Abdulghany-2017},
including undergraduate level textbooks \cite{CM,QM}.
Most of them deal with
the inertia tensor in explicit matrix representations.
In this paper, we employ Dirac's bra-ket 
notation~\cite{Dirac-1923,Dirac-1939,Dirac-1958} to define the
inertia tensor operator, from which one can project out the 
corresponding inertia tensor elements by applying a bra and a ket
for the unit basis vectors $\hat{\mathbf{q}}_i$ along the principal axes. 
While the completeness relation 
reflects the homogeneousness and isotropy of the Euclidean space
$\mathbf{1}=|\hat{\mathbf{q}}_i\rangle\langle\hat{\mathbf{q}}_i|$,
the inertia tensor operator acquires the anisotropy of the mass 
distribution so that each principal axis is distinct in general:
$\mathbf{I}=I_i|\hat{\mathbf{q}}_i\rangle\langle\hat{\mathbf{q}}_i|$.
The coordinate dependence of the inertia tensor of a system with the unit basis vector $\hat{\mathbf{e}}_i$ can be read off
directly from the operator by finding the projections:
$\mathbbm{I}=(I_{ij})=(\langle\hat{\mathbf{e}}_i|\mathbf{I}|\hat{\mathbf{e}}_j\rangle)$. The operator method is independent
of the coordinate system. Thus, the computation of the
matrix element is minimal because one needs to compute only 
three fundamental scalar integrals $I_i$ that are the principal moments.
An extension to the covariant approach as in \cite{Ee-2017} is straightforward
because of the coordinate independence of this formulation.

This paper is organized as follows: In Sec.~\ref{sec:def}, we list the definitions
of the quantities
involving the bra-ket notation and the inertia tensor operator. A summary of
the relationship between the principal axes and the coordinate dependence
of the inertia tensor is given in Sec.~\ref{sec:PA}. As applications of 
the approach developed in the preceding sections, we compute
the inertia tensor for an ellipsoid in the three- and $N$-dimensional
Euclidean spaces in Sec.~\ref{sec:I-Ellipsoid} and
we conclude in Sec.~\ref{sec:dis}. Appendices are attached
to list the basis operators for the rank-2 Cartesian tensors
in two and three dimensions.
\section{Definitions\label{sec:def}}
\subsection{Bra-Ket Notation}
By employing Dirac's bra-ket notation~\cite{Dirac-1923,Dirac-1939,Dirac-1958}, 
one can express
the scalar product of any two Euclidean
vectors $\mathbf{U}$ and $\mathbf{V}$ in $N$ dimensions as
\begin{equation}
\mathbf{U}\cdot\mathbf{V}=
\langle \mathbf{U}|\mathbf{V}\rangle=
\langle \mathbf{V}|\mathbf{U}\rangle.
\end{equation} 
Let us consider unit basis vectors $\hat{\mathbf{e}}_i$ in the Cartesian coordinate system
that satisfy the orthonormal condition and the completeness relation
\begin{eqnarray}
\langle\hat{\mathbf{e}}_j|\hat{\mathbf{e}}_k\rangle
&=&
\delta_{jk},
\\
\label{identity-op}
|\hat{\mathbf{e}}_i\rangle\langle\hat{\mathbf{e}}_i|
&=&\mathbf{1},
\end{eqnarray}
where $\delta_{jk}$ is the Kronecker delta and $i$ is summed over $1$, $\cdots$, $N$. In the remainder of this paper,
any repeated vector indices are assumed to be summed over unless specified.
Any vector can be expanded in terms of the bases as
\begin{equation}
\label{V-expansion}
|\mathbf{V}\rangle=V_i|\hat{\mathbf{e}}_i\rangle,
\end{equation} 
where the Cartesian component $V_i$
of $\mathbf{V}$ can be read off as
\begin{equation}
V_i=\langle \hat{\mathbf{e}}_i|\mathbf{V}\rangle.
\end{equation}
The identity operator $\mathbf{1}$ is invariant under rotation. 
Thus, the corresponding matrix representation $\mathbbm{1} $ is 
independent of the coordinate system:
\begin{equation}
\mathbbm{1}
=(\langle \hat{\mathbf{e}}_i|\mathbf{1}|\hat{\mathbf{e}}_{j} \rangle)
=(\langle \hat{\mathbf{e}}_i| \hat{\mathbf{e}}_{j} \rangle)
=(\delta_{ij}).
\end{equation}

\subsection{Inertia Tensor Operator}
The $ij$ element of the inertia tensor of a rigid body is a symmetric rank-2 Cartesian tensor:
\begin{equation}
\label{I-ij-element}
I_{ij}=\int d^3\mathbf{x}
\, \rho(\mathbf{x})
[
\delta_{ij}\mathbf{x}^2-x_ix_j
],
\end{equation}
 where $\rho(\mathbf{x})$ is the mass density at point $\mathbf{x}$ and the integration
is over the space. The rotational properties of the inertia tensor 
is completely determined by the scalar field $\rho(\textbf{x})$.
By employing Dirac's bra-ket notation, we can express the \textit{inertia tensor operator} as
\begin{equation}
\label{I-operator}
\mathbf{I}
=\int d^3\mathbf{x} \rho(\textbf{x})
\big[
\langle\textbf{x}|\textbf{x}\rangle \mathbf{1}-|\textbf{x}\rangle\langle\textbf{x}|\,
\big].
\end{equation}
 The matrix representation of the inertia tensor $\mathbbm{I}=(I_{ij})$ in the coordinate system with the basis vectors $\hat{\mathbf{e}}_i$'s 
can be expressed as
\begin{equation}
\label{Iij-bra-ket}
I_{ij}=\langle \hat{\mathbf{e}}_i|\mathbf{I}|\hat{\mathbf{e}}_j\rangle.
\end{equation}
 Note that the quantities like $V_i$ in Eq.~\eqref{V-expansion}
or $I_{ij}$ in Eq.~\eqref{I-ij-element} depend on the choice of basis vectors or
coordinate system. However, the operator representation as in 
Eq.~\eqref{I-operator} is independent of the coordinate system unless
it is rotationally invariant as in Eq.~\eqref{identity-op}.

\section{Geometry and Principal Axes\label{sec:PA}}
We consider an elliptic plate that has the simplest geometry without having 
rotational symmetry in two dimensions.
The principal axes and the corresponding principal values for the rigid body
are first considered,
not based on the inertia tensor
but based on the geometry only, through the parametric equation.
The operator for the geometry of the rigid body is extended to 
the inertia tensor operator.
\subsection{Principal Axes}
The parametric curve of an ellipse on the $xy$ plane is
\begin{equation}
\label{ellipse}
\frac{x^2}{a^2}+\frac{y^2}{b^2}=1,
\end{equation}
 where $a$ and $b$ are the semi-major and semi-minor axes
that are along the $x$ and $y$ axes, respectively.
An important feature of this expression is that the left side
is the sum of squared numbers. We call the coordinate axes 
of the system the 
\textit{principal axes}
 with which a physical observable can be expressed as the sum of
 squared numbers. In this case, those principal axes consist
of the major and the minor axes. Thus, any expression for an
observable acquires the most compact form. This is a reflection
of the symmetries of a physical observable.

We define $x_1\equiv x$, $x_2\equiv y$, and the orthonormal basis vectors
$\hat{\mathbf{q}}_i$'s along the principal axes to find that
\begin{equation}
|\mathbf{x}\rangle=x_i|\hat{\mathbf{q}}_i\rangle.
\end{equation}
 The parametric equation \eqref{ellipse} can be represented by
\begin{equation}
\langle \mathbf{x}|\mathbf{S}_{\textrm{P}}|\mathbf{x}\rangle=1,
\end{equation}
 where $\mathbf{S}_{\textrm{P}}$ is the tensor operator defined by 
\begin{equation}
\label{S-operator}
\mathbf{S}_{\textrm{P}}=
\alpha_{\textrm{P}}|\hat{\mathbf{q}}_1\rangle\langle\hat{\mathbf{q}}_1|
+
\beta_{\textrm{P}}|\hat{\mathbf{q}}_2\rangle\langle\hat{\mathbf{q}}_2|.
\end{equation}
 The operator $\mathbf{S}_{\textrm{P}}$ is a special case of $\gamma=0$
in a general symmetric tensor operator in Eq.~\eqref{S-2D}.
The subscript P indicates that the operator is written
in the frame whose coordinate axes are principal axes.
Here, the coefficients 
\begin{subequations}
\label{ab-prin}
\begin{eqnarray}
\alpha_{\textrm{P}}&=&\frac{1}{a^2},
\\
\beta_{\textrm{P}}&=&\frac{1}{b^2},
\end{eqnarray}
\end{subequations}
 are the \textit{principal values}
for the operator $\mathbf{S}_{\textrm{P}}$ 
corresponding to principal axes $\hat{\mathbf{q}}_1$ and $\hat{\mathbf{q}}_2$, respectively.

The operator for the inertia tensor for a two-dimensional rigid body must be of the form
\begin{equation}
\mathbf{I}=I_1|\hat{\mathbf{q}}_1\rangle\langle\hat{\mathbf{q}}_1|
+
I_2|\hat{\mathbf{q}}_2\rangle\langle\hat{\mathbf{q}}_2|.
\end{equation}
This is true for any two-dimensional rigid body
as well as the elliptic plate that we consider in this section. 
The reason is that we can always construct an equivalent elliptic plate that has the
same principal axes with the same principal values.
 
\subsection{Non-principal Axes}
\begin{figure}
\begin{center}
\includegraphics[width=50mm]{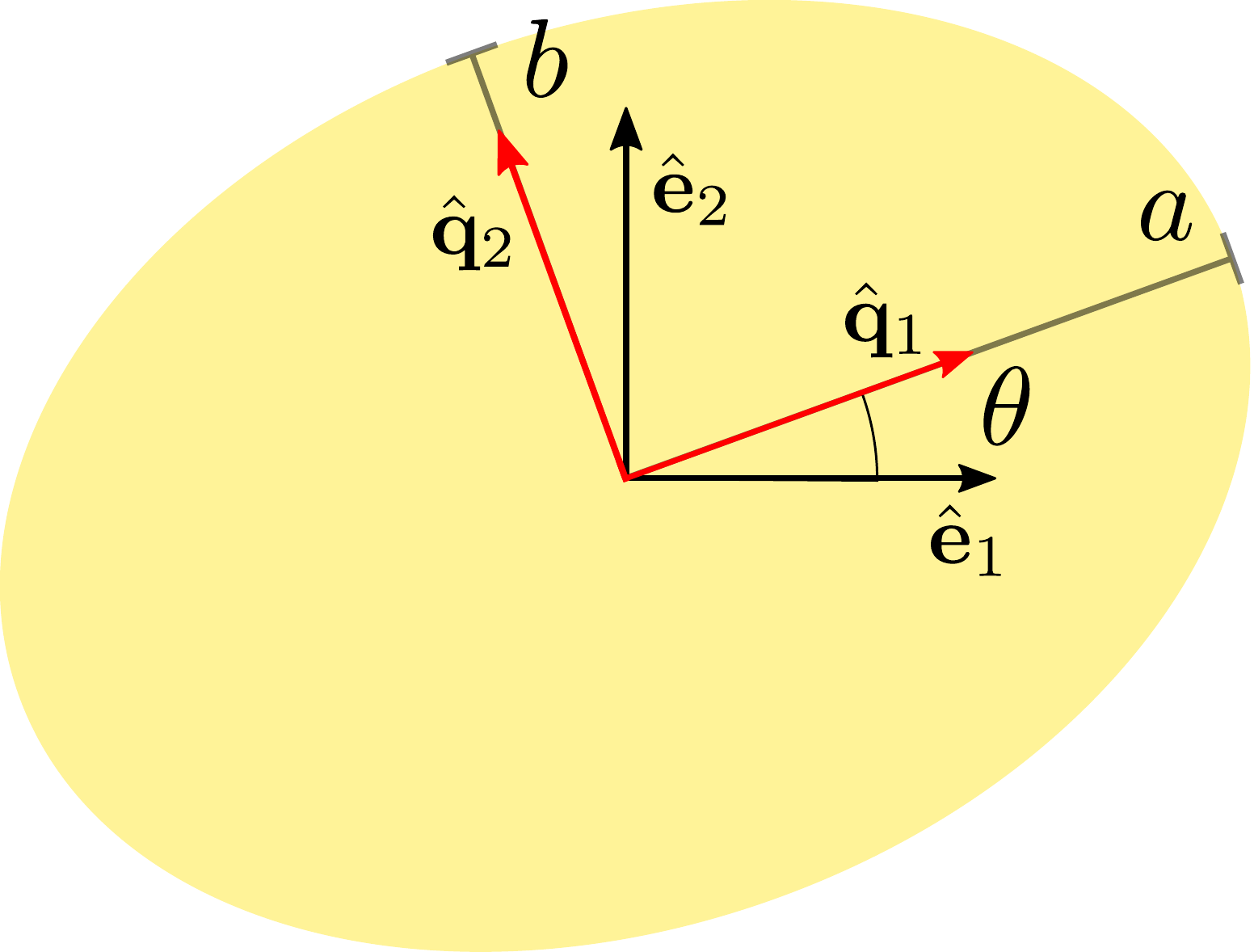}
\caption{
An elliptic plate with semi-major axis $a$ and semi-minor axis $b$. 
The principal axes $\hat{\mathbf{q}}_i$'s of the plate 
are related to $\hat{\mathbf{e}}_i$'s under rotation by an angle $\theta$.  
\label{fig:2D-ep}
}
\end{center}
\end{figure}
We shall find that the operator $\mathbf{S}_{\textrm{P}}$ in Eq.~\eqref{S-operator}
acquires additional contributions in a frame of reference
whose axes are not principal.
We consider another set of orthonormal basis vectors $\hat{\mathbf{e}}_i$'s 
that are related to the principal axes under rotation as shown in Fig.~\ref{fig:2D-ep}: 
\begin{eqnarray}
|\hat{\mathbf{q}}_1\rangle&=&\phantom{-}
\cos\theta|\hat{\mathbf{e}}_1\rangle+\sin\theta|\hat{\mathbf{e}}_2\rangle,
\\
|\hat{\mathbf{q}}_2\rangle&=&
-\sin\theta|\hat{\mathbf{e}}_1\rangle+\cos\theta|\hat{\mathbf{e}}_2\rangle.
\end{eqnarray}
The corresponding directional cosines are given by
\begin{eqnarray}
\begin{matrix}
\langle\hat{\mathbf{e}}_1|\hat{\mathbf{q}}_1\rangle=\phantom{+}\cos\theta,&&
\langle\hat{\mathbf{e}}_2|\hat{\mathbf{q}}_1\rangle=\sin\theta,
\nonumber\\
\langle\hat{\mathbf{e}}_1|\hat{\mathbf{q}}_2\rangle=-\sin\theta,&&
\langle\hat{\mathbf{e}}_2|\hat{\mathbf{q}}_2\rangle=\cos\theta.
\end{matrix}
\end{eqnarray}
By multiplying the identity operators in Eq.~\eqref{identity-op} to both sides of Eq.~\eqref{S-operator}, 
we can express the operator $\mathbf{S}_{\textrm{P}}$ 
in terms of the basis kets $|\hat{\mathbf{e}}_i\rangle$'s as 
\begin{eqnarray}
\label{S-ellipse}
\mathbf{S}_{\textrm{P}}
&=&
\mathbf{1}
\,
\mathbf{S}_{\textrm{P}}
\mathbf{1}
\nonumber\\
&=&
\Big(
\alpha_{\textrm{P}}\langle\hat{\mathbf{e}}_i|\hat{\mathbf{q}}_1\rangle 
\langle\hat{\mathbf{q}}_1|\hat{\mathbf{e}}_j\rangle
+
\beta_{\textrm{P}}\langle\hat{\mathbf{e}}_i|\hat{\mathbf{q}}_2\rangle 
\langle\hat{\mathbf{q}}_2|\hat{\mathbf{e}}_j\rangle
\Big)
|\hat{\mathbf{e}}_i\rangle \langle\hat{\mathbf{e}}_j|
\nonumber\\
&=&
\tfrac{1}{2}(\alpha_{\textrm{P}}+\beta_{\textrm{P}})\mathbf{1}
+ \tfrac{1}{2} (\alpha_{\textrm{P}}-\beta_{\textrm{P}})(\mathbf{S}_1 \cos2\theta +\mathbf{S}_2 \sin2\theta )
\nonumber\\
&\equiv&
s_0\mathbf{1}
+
s_1\mathbf{S}_1
+s_2
\mathbf{S}_2, 
\end{eqnarray}
where 
the operators $\mathbf{S}_i$'s are defined in Eqs.~\eqref{def-s0}, \eqref{def-s1}, 
and \eqref{def-s2}
and
\begin{subequations}
\begin{eqnarray}
	s_0
	&=&
	\tfrac{1}{2}(\alpha_{\textrm{P}}+\beta_{\textrm{P}}),
\\
	s_1
	&=&
	\tfrac{1}{2}(\alpha_{\textrm{P}}-\beta_{\textrm{P}})\cos2\theta,
\\
	s_2
	&=&
	\tfrac{1}{2}(\alpha_{\textrm{P}}-\beta_{\textrm{P}})\sin2\theta.
\end{eqnarray} 
\end{subequations}
The off-diagonal contribution emerges 
through the contribution proportional to $\mathbf{S}_2$.
Note that the parametrization in Eq.~\eqref{S-ellipse} is applicable to
any real symmetric rank-2 tensor operator.

Additionally, $s_0$ and $s_1^2+s_2^2$ are invariant separately:
\begin{eqnarray}
\sqrt{s_1^2+s_2^2} 
&=& \tfrac{1}{2}|\alpha_{\textrm{P}}-\beta_{\textrm{P}}|=\tfrac{1}{2}\epsilon(\alpha_{\textrm{P}}-\beta_{\textrm{P}}),
\\
s_0^2+s_1^2 +s_2^2
&=&
\tfrac{1}{2}(\alpha^2_{\textrm{P}}+\beta^2_{\textrm{P}}),
\end{eqnarray}
 where $\epsilon$ is the sign of $\alpha_{\textrm{P}}-\beta_{\textrm{P}}$:
\begin{equation}
\epsilon=\textrm{sgn}[\alpha_{\textrm{P}}-\beta_{\textrm{P}}].
\end{equation}
 The angle $\theta$ can be determined as
\begin{eqnarray}
\cos2\theta&=&\frac{\epsilon s_1}{\sqrt{s_1^2+s_2^2}},
\\
\sin2\theta&=& \frac{\epsilon s_2}{\sqrt{s_1^2+s_2^2}},
\\
\theta&=& \frac{1}{2}\arctan\frac{s_2}{s_1}.
\end{eqnarray}
The principal values can be expressed as
\begin{eqnarray}
\alpha_{\textrm{P}}&=&s_0+\epsilon\sqrt{s_1^2+s_2^2},
\\
\beta_{\textrm{P}}&=&s_0-\epsilon\sqrt{s_1^2+s_2^2},
\end{eqnarray}
which correspond, respectively, to the principal axes
$\hat{\mathbf{q}}_1$ and $\hat{\mathbf{q}}_2$.

\subsection{Diagonalization}
We determine the angle to diagonalize 
a real symmetric rank-2 tensor operator $\mathbf{S}$ in Eq.~\eqref{S-2D},
\begin{eqnarray}
\label{S-2D2}
\mathbf{S} 
=s_0\mathbf{1}+s_1\mathbf{S}_1+s_2\mathbf{S}_2,
\end{eqnarray}
 to reach the form in Eq.~\eqref{S-operator}. In most textbooks, the readers are advised to solve the eigenvalue problem
to find the principal axes.
Although such an approach is mathematically straightforward, it does not tend to provide 
students with physical intuitions. Thus, one must 
investigate the solution further to interpret the physical meaning even after obtaining the result.
Here, we rotate the orthonormal basis vectors from
 $\{\hat{\mathbf{e}}_1 ,\hat{\mathbf{e}}_2 \}$
to
 $\{\hat{\mathbf{q}}_1 ,\hat{\mathbf{q}}_2 \}$ as shown in Fig.~\ref{fig:2D-ep} as
\begin{eqnarray}
|\hat{\mathbf{e}}_1\rangle&=&
\cos\theta|\hat{\mathbf{q}}_1\rangle-\sin\theta|\hat{\mathbf{q}}_2\rangle,
\\
|\hat{\mathbf{e}}_2\rangle&=&
\sin\theta|\hat{\mathbf{q}}_1\rangle+\cos\theta|\hat{\mathbf{q}}_2\rangle.
\end{eqnarray}
 Substituting these into Eq.~\eqref{S-2D2}, we obtain 
\begin{eqnarray}
\mathbf{S}
&=&
(s_0+s_1\cos2\theta+s_2\sin2\theta)|\hat{\mathbf{q}}_1\rangle\langle\hat{\mathbf{q}}_1|
\nonumber\\&&
+
(s_0-s_1\cos2\theta-s_2\sin2\theta)|\hat{\mathbf{q}}_2\rangle\langle\hat{\mathbf{q}}_2|
\nonumber\\
&&+
(-s_1\sin2\theta+s_2\cos2\theta)(|\hat{\mathbf{q}}_1\rangle\langle\hat{\mathbf{q}}_2|+|\hat{\mathbf{q}}_2\rangle\langle\hat{\mathbf{q}}_1|).\phantom{X}
\end{eqnarray}

The operator $\mathbf{S}$ reaches the principal form in which the crossing terms
$|\hat{\mathbf{q}}_1\rangle\langle\hat{\mathbf{q}}_2|$ and 
$|\hat{\mathbf{q}}_2\rangle\langle\hat{\mathbf{q}}_1|$ vanish: 
\begin{equation}
\mathbf{S}=\alpha_{\textrm{P}}|\hat{\mathbf{q}}_1\rangle\langle \hat{\mathbf{q}}_1|+\beta_{\textrm{P}}|\hat{\mathbf{q}}_2\rangle\langle \hat{\mathbf{q}}_2|,
\end{equation}
 where $\alpha_{\textrm{P}}$ and $\beta_{\textrm{P}}$ are principal values in Eq.~\eqref{ab-prin}.
The vanishing condition is satisfied when
\begin{equation}
\label{eq:2D-vanish}
s_1\sin2\theta_{\textrm{P}}=s_2\cos2\theta_{\textrm{P}}.
\end{equation}
The angle $\theta_{\textrm{P}}$ can 
always be chosen as
\begin{equation}
\label{eq:angle-choice}
\theta_{\textrm{P}}=\frac{1}{2} \arctan\frac{s_2}{s_1},
\quad
0\leq\theta_{\textrm{P}}\leq \pi/2.
\end{equation}
A convenient choice of the solution set is 
\begin{equation}
\label{e2theta}
 (\cos2\theta_{\textrm{P}},\sin2\theta_{\textrm{P}})=\frac{1}{\sqrt{s_1^2+s_2^2}} (s_1,s_2) .
\end{equation}
 The principal values are determined as
\begin{eqnarray}
\alpha_{\textrm{P}}&=&
s_0+s_1\cos2\theta_{\textrm{P}}+s_2\sin2\theta_{\textrm{P}}
\nonumber\\
&=&
s_0+\sqrt{s_1^2+s_2^2},
\\
\beta_{\textrm{P}}&=&
s_0-s_1\cos2\theta_{\textrm{P}}-s_2\sin2\theta_{\textrm{P}}
\nonumber\\
&=&s_0-\sqrt{s_1^2+s_2^2},
\end{eqnarray}
 where we have made use of the solution set in Eq.~\eqref{e2theta}.

By making use of the trigonometric identity $\cos^2\theta-\sin^2\theta=\cos2\theta$
and applying the choice in Eq.~\eqref{eq:angle-choice},
we can always determine $\cos\theta_{\text P}$ and $\sin\theta_{\text P}$ uniquely as
\begin{subequations}
\label{eq:2D-cos-sin}
\begin{eqnarray}
\cos \theta_{\textrm{P}}&=&\sqrt{\tfrac{1}{2}(1+\cos 2\theta_{\textrm{P}})}
\nonumber\\
&=&\sqrt{\frac{1}{2}\left[1+\frac{s_1}{\sqrt{s_1^2+s_2^2}}\right]},
\\
\sin \theta_{\textrm{P}}&=&\sqrt{\tfrac{1}{2}(1-\cos 2\theta_{\textrm{P}})}
\nonumber\\
&=&\sqrt{\frac{1}{2}\left[1-\frac{s_1}{\sqrt{s_1^2+s_2^2}}\right]}.
\end{eqnarray}
\end{subequations}
The corresponding principal axes are
\begin{eqnarray}
\label{eq:2D-trans}
|\hat{\mathbf{q}}_1\rangle&=&\phantom{-}
\cos\theta_{\textrm{P}}|\hat{\mathbf{e}}_1\rangle+\sin\theta_{\textrm{P}}|\hat{\mathbf{e}}_2\rangle,
\\
|\hat{\mathbf{q}}_2\rangle&=&
-\sin\theta_{\textrm{P}}|\hat{\mathbf{e}}_1\rangle+\cos\theta_{\textrm{P}}|\hat{\mathbf{e}}_2\rangle.
\end{eqnarray}

Our determination of the rotation matrix that diagonalizes 
the symmetric tensor operator is manifest. This approach is
particularly convenient and intuitive in determining the eigenvector.

\section{Inertia Tensor of an Ellipsoid\label{sec:I-Ellipsoid}}
In Sec.~\ref{sec:PA}, we  found that the parametric equation
for the ellipse can be expressed minimally by choosing 
the principal axes as the coordinate axes. The same principal axes
can be used to express the inertia tensor that shares the principal axes
because it is a fundamental property of symmetric tensor operators.
In this section, we compute the inertia tensor operators for uniform ellipsoids in the three- and $N$-dimensional Euclidean spaces.
Because the inertia tensor operator for the elliptic plate 
can be obtained from the three-dimensional counterpart,
we omit that calculation. 
\subsection{Ellipsoid in Three Dimensions}
As an application, we compute the inertia tensor for a uniform ellipsoid of mass $M$ in  three-dimensional Euclidean space. 
The parametric equation \eqref{ellipse} can be generalized to $\sum_{i=1}^3x_i^2/a_i^2=1$. Then the mass distribution is
\begin{equation}
\label{eq:3D-rho}
\rho(\textbf{x},\mathbf{q}_1,\mathbf{q}_2,\mathbf{q}_3)=\frac{3M}{4\pi a_1a_2a_3}
\Theta[1-\langle\textbf{x}|\mathbf{E}|\textbf{x}\rangle ],
\end{equation}
 where $\Theta$ is the Heaviside step function and $\mathbf{E}$ is a symmetric tensor operator
\begin{equation}
\mathbf{E}
=
 \frac{1}{a_i^2}| \hat{\mathbf{q}}_i\rangle\langle \hat{\mathbf{q}}_i|.
\end{equation}
Here, $\hat{\mathbf{q}}_i$ are the orthonormal basis vectors along the principal axes of the ellipsoid.
Because a real symmetric rank-2 tensor operator in three dimensions can be expressed as 
Eq.~\eqref{SymmetricOperatorExpansion}, the inertia tensor operator must be expanded as 
\begin{eqnarray}
\mathbf{I}
&=&\int d^3\mathbf{x} 
\,
\rho(\textbf{x},\mathbf{q}_1,\mathbf{q}_2,\mathbf{q}_3)
\big[
\langle\textbf{x}|\textbf{x}\rangle \mathbf{1}-|\textbf{x}\rangle\langle\textbf{x}|\,
\big]
\nonumber\\
&=&
\alpha_i | \hat{\mathbf{q}}_i\rangle\langle \hat{\mathbf{q}}_i|
+\sum_{i<j}\gamma_{ij} \mathbf{S}_{ij},
\label{eq:3-dim-iten}
\end{eqnarray}
where 
$\mathbf{S}_{ij}=
| \hat{\mathbf{q}}_i\rangle\langle \hat{\mathbf{q}}_j|+
| \hat{\mathbf{q}}_j\rangle\langle \hat{\mathbf{q}}_i|$.
The scalar factors in Eq.~\eqref{eq:3-dim-iten} can be read off as
\begin{eqnarray}
\alpha_i&=&
\langle \hat{\mathbf{q}}_i|\mathbf{I}|\hat{\mathbf{q}}_i\rangle,
\\
\gamma_{ij}&=& \langle \hat{\mathbf{q}}_i|\mathbf{I}|\hat{\mathbf{q}}_j\rangle.
\end{eqnarray}
 Therefore, the matrix element 
$I_{ij}=\langle \hat{\mathbf{e}}_i|\mathbf{I}|\hat{\mathbf{e}}_j\rangle $
 of the inertia tensor
$\mathbbm{I}$ 
can be determined
as
\begin{equation}
\label{eq:3D-inertia}
I_{ij}=
 \alpha_k q_{ki}q_{kj}
+\sum_{k<\ell}\gamma_{k\ell} \left(q_{ki}q_{\ell j}+q_{\ell i}q_{kj}\right), 
\end{equation}
 where $q_{ki}\equiv\langle\hat{\mathbf{q}}_k|\hat{\mathbf{e}}_i\rangle=%
\hat{\mathbf{q}}_k\cdot \hat{\mathbf{e}}_i$. 

Let us consider the scalar integrals for $\alpha_{i}$ and $\gamma_{ij}$:
\begin{eqnarray}
\alpha_i&=& \frac{3M}{4\pi a_1a_2a_3}
\int_V d^3\mathbf{x}  \,
(\langle\textbf{x}|\textbf{x}\rangle-\langle\textbf{x}|\hat{\mathbf{q}}_i\rangle^2) ,
\nonumber\\
\gamma_{ij}
&=&-\frac{3M}{4\pi a_1a_2a_3}
\int_V d^3\mathbf{x}
 \langle\hat{\mathbf{q}}_i|\textbf{x}\rangle\langle\textbf{x}|\hat{\mathbf{q}}_j\rangle,
\label{eq:int-alp-gam}
\end{eqnarray}
where 
\begin{equation}
\int_V d^3\mathbf{x}
\equiv
\int d^3\mathbf{x}\,\Theta[1-\langle\textbf{x}|\mathbf{E}|\textbf{x}\rangle ].
\end{equation}
Here, we denote $V$ by the region inside the ellipsoid. If we define 
\begin{equation}
\label{y-def}
\mathbf{y}=\frac{\langle \hat{\mathbf{q}}_i|\mathbf{x}\rangle}{a_i}\hat{\mathbf{q}}_i,
\end{equation}
 then the region $V$ is 
parametrized by 
\begin{equation}
\langle \mathbf{y} |\mathbf{y}\rangle\le 1.
\end{equation}
 In addition, we find that
\begin{subequations}
\label{eq:COV-3dim}
\begin{eqnarray}
\int_V d^3\mathbf{x} 
&=&a_1a_2a_3
\int_V d^3\mathbf{y},
\\ 
\langle\textbf{x}|\textbf{x}\rangle&=&
\langle\textbf{x}
|\hat{\mathbf{q}}_i\rangle\langle \hat{\mathbf{q}}_i|
\textbf{x}\rangle
 =
a_k^2\langle\textbf{y}|\hat{\mathbf{q}}_k\rangle^2,
\\ 
\langle\textbf{x}|\hat{\mathbf{q}}_i\rangle^2
&=&
a_i^2\langle\textbf{y}|\hat{\mathbf{q}}_i\rangle^2,
\end{eqnarray}
\end{subequations}
where $i$ in the last line is not summed over.

Making a change of variable in Eq.~\eqref{eq:int-alp-gam} by using the identities in Eq.~\eqref{eq:COV-3dim} 
and carrying out the integration, we obtain
\begin{eqnarray}
\alpha_i&=& \frac{3M}{4\pi a_1a_2a_3}
\int_V d^3\mathbf{x}  \,
(\langle\textbf{x}|\textbf{x}\rangle-\langle\textbf{x}|\hat{\mathbf{q}}_i\rangle^2) 
\nonumber\\
&=&\frac{3M}{4\pi }a_k^2(1-\delta_{ki}) 
\int_V d^3\mathbf{y}  \,
 \langle\textbf{y}|\hat{\mathbf{q}}_k\rangle^2,
\nonumber\\
\gamma_{ij}
&=&-\frac{3M}{4\pi a_1a_2a_3}
\int_V d^3\mathbf{x}
 \langle\hat{\mathbf{q}}_i|\textbf{x}\rangle\langle\textbf{x}|\hat{\mathbf{q}}_j\rangle
\nonumber\\
&=&0,
\label{eq:3D-scal}
\end{eqnarray}
where 
$\gamma_{ij}=0$ because the integrand is an odd function of both 
$\langle\textbf{x}|\hat{\mathbf{q}}_i\rangle$ and
$\langle\textbf{x}|\hat{\mathbf{q}}_j\rangle$ while the region $V$ is even in those variables.
Note that the factor $(1-\delta_{ki})$ in $\alpha_i$ is the projection operator for the direction perpendicular to $\hat{\mathbf{q}}_i$. 
The isotropy of the Euclidean space guarantees that
\begin{eqnarray}
\int_V d^3\mathbf{y}  \,
 \langle\textbf{y}|\hat{\mathbf{q}}_k\rangle^2
 &=&\frac{1}{3}\int_V d^3\mathbf{y}  \,
 \langle\textbf{y}| \mathbf{y} \rangle 
\nonumber\\
 &=&\frac{4\pi}{3}\int_0^1 dr\,r^4
\nonumber\\
 &=&\frac{4\pi}{15}.
\end{eqnarray}
As a result, the principal moments are determined as
\begin{eqnarray}
\alpha_1&=&\frac{1}{5}M(a_2^2+a_3^2),
\nonumber\\
\alpha_2&=&\frac{1}{5}M(a_3^2+a_1^2),
\nonumber\\
\alpha_3&=&\frac{1}{5}M(a_1^2+a_2^2).
\label{eq:3dim-alpha}
\end{eqnarray}

\subsection{Ellipsoid in $N$ Dimensions}
As a pedagogical example, we evaluate the inertia tensor for a uniform ellipsoid of mass $M$ in $N$-dimensional Euclidean space.
The parametric equation of the ellipsoid in $N$ dimensions is
\begin{equation}
\sum_{i=1}^N\frac{x_i^2}{a_i^2}=1.
\end{equation}
 The volume of the ellipsoid is given by
\begin{equation}
\label{eq:ND-vol1}
V_N=\int d^N\mathbf{x} \, \Theta[1-\langle\textbf{x}|\mathbf{E}_N|\textbf{x}\rangle ],
\end{equation}
where $\mathbf{E}_N$ is a symmetric tensor operator
\begin{equation}
\mathbf{E}_N=
  \sum_{i=1}^N| {\mathbf{q}}_i\rangle\langle {\mathbf{q}}_i|
=\sum_{i=1}^N
 \frac{1}{a_i^2}| \hat{\mathbf{q}}_i\rangle\langle \hat{\mathbf{q}}_i|.
\end{equation}
Note that $\hat{\mathbf{q}}_i$ are the orthonormal basis vectors along the principal axes of the ellipsoid.
Introducing the rescaled vector $\mathbf{y}$ which is the $N$-dimensional
generalization of Eq.~\eqref{y-def}, we find that
\begin{eqnarray}
V_N
&=&\left(\prod_{i=1}^N a_i\right)
\int d^N\mathbf{y}\, \Theta[1- \mathbf{y}^2 ]
\nonumber\\
&=&\left(\prod_{i=1}^N a_i\right)
\Omega_N\int_0^1 dr\,r^{N-1},
\nonumber\\
&=&\frac{ \Omega_N}{N}\prod_{i=1}^N a_i,
\label{eq:ND-vol2}
\end{eqnarray}
 where $\Omega_N$ is the $N$-dimensional solid angle \cite{Ee-2017}
\begin{eqnarray}
\Omega_N=\frac{2\pi^{\tfrac{1}{2}N}}{\Gamma(\tfrac{1}{2}N)}.
\end{eqnarray}
 The mass distribution is
\begin{equation}
\label{eq:ND-mass}
\rho(\textbf{x},\mathbf{q}_1,\cdots,\mathbf{q}_N)=\frac{NM}{\Omega_N \prod_{i=1}^N a_i }
\Theta[1-\langle\textbf{x}|\mathbf{E}_N|\textbf{x}\rangle ].
\end{equation}
Because the inertia tensor operator is symmetric, it must be expanded as
\begin{eqnarray}
\mathbf{I}
&=&\int d^N\mathbf{x} 
\,
\rho(\textbf{x},\mathbf{q}_1,\cdots,\mathbf{q}_N)
\big[
\langle\textbf{x}|\textbf{x}\rangle \mathbf{1}-|\textbf{x}\rangle\langle\textbf{x}|\,
\big]
\nonumber\\
&=&
\sum_{i=1}^N\alpha_i | \hat{\mathbf{q}}_i\rangle\langle \hat{\mathbf{q}}_i|
+\sum_{i<j}\gamma_{ij} \mathbf{S}_{ij},
\label{eq:n-dim-Iten}
\end{eqnarray}
where 
$\mathbf{S}_{ij}=
| \hat{\mathbf{q}}_i\rangle\langle \hat{\mathbf{q}}_j|+
| \hat{\mathbf{q}}_j\rangle\langle \hat{\mathbf{q}}_i|.$

The scalar factors in Eq.~\eqref{eq:n-dim-Iten} can be determined as
\begin{eqnarray}
\alpha_i&=&
\langle \hat{\mathbf{q}}_i|\mathbf{I}|\hat{\mathbf{q}}_i\rangle,
\\
\gamma_{ij}&=& \langle \hat{\mathbf{q}}_i|\mathbf{I}|\hat{\mathbf{q}}_j\rangle.
\end{eqnarray}
Therefore, the matrix element 
$
I_{ij}=\langle \hat{\mathbf{e}}_i|\mathbf{I}|\hat{\mathbf{e}}_j\rangle 
$
 of the inertia tensor
$\mathbbm{I}$ can be computed
as
\begin{equation}
\label{eq:ND-inertia}
I_{ij}=
\sum_{k=1}^N\alpha_k q_{ki}q_{kj}
+\sum_{k<\ell}\gamma_{k\ell} \left(q_{ki}q_{\ell j}q_{\ell i}q_{kj}\right),
\end{equation}
where $q_{ki}\equiv\langle\hat{\mathbf{q}}_k|\hat{\mathbf{e}}_i\rangle=%
\hat{\mathbf{q}}_k\cdot \hat{\mathbf{e}}_i$. 

We evaluate the scalar integrals for $\alpha_i$ and $\gamma_{ij}$ as
\begin{eqnarray}
\label{eq:ND-scal}
\alpha_i&=& \frac{NM}{\Omega_N\prod_{i=1}^N a_i}
\int_V d^N\mathbf{x}  \,
(\langle\textbf{x}|\textbf{x}\rangle-\langle\textbf{x}|\hat{\mathbf{q}}_i\rangle^2) 
\nonumber\\
&=&\frac{NM}{\Omega_N }a_k^2(1-\delta_{ki}) 
\int_V d^{N}\mathbf{y}  \,
 \langle\textbf{y}|\hat{\mathbf{q}}_k\rangle^2,
\\
\gamma_{ij}
&=&-\frac{NM}{\Omega_N \prod_{i=1}^N a_i }
\int_V d^N\mathbf{x}
 \langle\hat{\mathbf{q}}_i|\textbf{x}\rangle\langle\textbf{x}|\hat{\mathbf{q}}_j\rangle
\nonumber\\
&=&0,
\end{eqnarray}
where $\gamma_{ij}=0$ because both 
$\langle\textbf{x}|\hat{\mathbf{q}}_i\rangle$ and
$\langle\textbf{x}|\hat{\mathbf{q}}_j\rangle$ in
 the integrand are odd while the region $V$ is even in those variables.
We notice that $(1-\delta_{ki})$ is the projection operator for the direction perpendicular to $\hat{\mathbf{q}}_i$.
By making use of the isotropy of the Euclidean space,
we obtain 
\begin{eqnarray}
\int_V d^N\mathbf{y}  \,
 \langle\textbf{y}|\hat{\mathbf{q}}_k\rangle^2
 &=&\frac{1}{N}\int_V d^N \mathbf{y}  \,
 \langle\textbf{y}| \mathbf{y} \rangle 
\nonumber\\
 &=&\frac{\Omega_N}{N}\int_0^1 dr\,r^{N+1}
\nonumber\\
 &=&\frac{\Omega_N}{N(N+2)}.
\end{eqnarray}
Therefore, 
\begin{equation}
\alpha_i=\frac{M}{N+2}\sum_{k\neq i}a_k^2,
\end{equation}
which reproduces the result in Eq.~\eqref{eq:3dim-alpha} when $N=3$.

\section{Discussion\label{sec:dis}}

In this paper, we have employed Dirac's bra-ket 
notation~\cite{Dirac-1923,Dirac-1939,Dirac-1958} to define
the inertia tensor operator $\mathbf{I}$ in Eq.~\eqref{I-operator}.
This operator is independent of the choice of bases or coordinate system. 
The matrix elements $I_{ij}$ of the inertia tensor can be read off by applying the 
bra and the ket for the unit basis vectors $\hat{\mathbf{e}}_i$ as
$I_{ij}=\langle\hat{\mathbf{e}}_i|\mathbf{I}|\hat{\mathbf{e}}_j\rangle$.
The computation of the
matrix element is minimal when the principal axes are taken to be the bases 
because one needs only to compute the
three fundamental scalar integrals $I_i$ that are the principal moments.

We have considered an elliptic plate, which has the simplest geometry without having 
rotational symmetry in two dimensions.
The principal axes and the corresponding principal values for the elliptic plate
have been determined only based on the geometry through 
the introduction of the symmetric operator
that satisfies the parametric equation of the ellipse.
It is worthwhile to mention that the inertia tensor operator of any planar lamina can always be represented by
the symmetric tensor of the elliptic plate that has the same principal values.
By making use of a general symmetric tensor operator in Eq.~\eqref{S-2D}, 
we have constructed the method of diagonalization without solving an eigenvalue problem.
This approach is particularly convenient and intuitive in determining the eigenvector.

As applications, we have determined the inertia tensor operators 
for uniform ellipsoids in three and $N$ dimensions.
The bra-ket approach allows one to express the tensor operator in a compact form,
and one has to evaluate only the minimal number of the scalar integrals, 
the principal moments of inertia. 
We have demonstrated that 
the bra-ket notation interplayed with various rescaling of the state kets 
greatly simplifies the evaluation of the scalar integrals corresponding to 
the principal moments of the $N$-dimensional ellipsoid. 

The bra-ket approach, which is independent of the coordinate choice is naturally extended to the covariant approach that is given, for example, in \cite{Ee-2017}. The exploitation of the bra-ket notation to compute the inertia tensor in classical mechanics 
should provide undergraduate students with 
a strong background 
that may result in a smooth transition to attack abstract quantum mechanical problems.

\begin{acknowledgments}
As members of the Korea Pragmatist Organization for Physics Education
(\textsl{KPOP}$\mathscr{E}$), 
the authors thank the remaining members of \textsl{KPOP}$\mathscr{E}$ 
for useful discussions.
This work is supported in part by the National Research Foundation
of Korea (NRF) grant funded by the Korea government (MSIT) 
under Contract Nos.\,NRF-2020R1A2C3009918 (U-R.K. and J.L.),
NRF-2017R1E1A1A01074699 (J.L.), 
 and NRF-2019R1A6A3A01096460 (U-R.K.), 
and the BK21+ program at Korea University, Initiative for Creative and Independent Scientists.
\end{acknowledgments}
\appendix
\section{Basis Tensor Operator in Two Dimensions}
In two dimensions, at most two linearly independent vectors are available.
By making use of the Gram-Schmidt orthogonalization procedure,
we can always express the operator in terms of two orthogonal vectors $\mathbf{a}$ and $\mathbf{b}$.
We set $\hat{\mathbf{e}}_1$ and $\hat{\mathbf{e}}_2$ to be the orthonormal basis vectors along these
directions.
The real symmetric tensor operator must be of the form
\begin{eqnarray}
\label{S-2D}
\mathbf{S} 
&=&
\alpha|\hat{\mathbf{e}}_1\rangle\langle\hat{\mathbf{e}}_1|+
\beta|\hat{\mathbf{e}}_2\rangle\langle\hat{\mathbf{e}}_2|
+\gamma(|\hat{\mathbf{e}}_1\rangle\langle\hat{\mathbf{e}}_2|+
|\hat{\mathbf{e}}_2\rangle\langle\hat{\mathbf{e}}_1|)
\nonumber\\
&=&
s_i\mathbf{S}_i,
\end{eqnarray}
 where $s_i$ is a scalar 
\begin{eqnarray}
\label{s0-2D}
s_0
&=&\frac{1}{2}(\alpha+\beta),
\\
\label{s1-2D}
s_1
&=&
\frac{1}{2}(\alpha-\beta),
\\
\label{s2-2D}
s_2
&=&
\gamma.
\end{eqnarray}
Also, the operators $\mathbf{S}_i$ are defined by
\begin{eqnarray}
\label{def-s0}
\mathbf{S}_0
&=&
|\hat{\mathbf{e}}_1\rangle\langle\hat{\mathbf{e}}_1|+
|\hat{\mathbf{e}}_2\rangle\langle\hat{\mathbf{e}}_2|=\mathbf{1},
\\
\label{def-s1}
\mathbf{S}_1
&=&
|\hat{\mathbf{e}}_1\rangle\langle\hat{\mathbf{e}}_1|-
|\hat{\mathbf{e}}_2\rangle\langle\hat{\mathbf{e}}_2|,
\\
\label{def-s2}
\mathbf{S}_2
&=&
|\hat{\mathbf{e}}_1\rangle\langle\hat{\mathbf{e}}_2|+
|\hat{\mathbf{e}}_2\rangle\langle\hat{\mathbf{e}}_1|.
\end{eqnarray}
 From these definitions, we find that
\begin{eqnarray}
\label{Se1e2Relation}
|\hat{\mathbf{e}}_1\rangle\langle\hat{\mathbf{e}}_1|&=&\frac{1}{2}(\mathbf{S}_0+\mathbf{S}_1),
\\
|\hat{\mathbf{e}}_2\rangle\langle\hat{\mathbf{e}}_2|&=&\frac{1}{2}(\mathbf{S}_0-\mathbf{S}_1). 
\end{eqnarray}
These basis tensor operators $\mathbf{S}_i$'s have a common feature:
\begin{equation}
\mathbf{S}_i^2=\mathbf{1},
\end{equation}
 where $i=0$, 1, 2 and the vector index $i$ is not summed over.

Their matrix representations 
$\mathbbm{S}_k\equiv (\langle \hat{\mathbf{e}}_{i}| \mathbf{S}_{k}|\hat{\mathbf{e}}_{j}\rangle)$ are
\begin{eqnarray}
\mathbbm{S}_0
&=&
\begin{pmatrix}
1&\phantom{+}0\\
0&\phantom{+}1\end{pmatrix},
\\
\mathbbm{S}_1
&=&
\begin{pmatrix}
1&\phantom{+}0
\\
0&-1\end{pmatrix},
\\
\mathbbm{S}_2
&=&
\begin{pmatrix}
0&\phantom{+}1
\\
1&\phantom{+}0\end{pmatrix}.
\end{eqnarray}
Note that only $\mathbbm{S}_0$ is traceful, and
that the remaining matrices $\mathbbm{S}_1$ and $\mathbbm{S}_2$ are both symmetric and traceless.
The last two matrices are identical to the Pauli sigma matrices: 
$\mathbbm{S}_1=\bbsigma_3$ and $\mathbbm{S}_2=\bbsigma_1$.

\section{Basis Tensor Operator in Three Dimensions}
In three dimensions, at most three linearly independent vectors are available.
 By making use of the Gram-Schmidt orthogonalization procedure,
we can always express the operator in terms of three orthogonal vectors $\hat{\mathbf{e}}_1$, 
$\hat{\mathbf{e}}_2$, and
$\hat{\mathbf{e}}_3$.
The real symmetric tensor operator must be of the form
\begin{eqnarray}
\label{SymmetricOperatorExpansion}
\mathbf{S} 
&=&
\alpha|\hat{\mathbf{e}}_1\rangle\langle\hat{\mathbf{e}}_1|
+\beta|\hat{\mathbf{e}}_2\rangle\langle\hat{\mathbf{e}}_2|
+\gamma|\hat{\mathbf{e}}_3\rangle\langle\hat{\mathbf{e}}_3|
+\mu(|\hat{\mathbf{e}}_1\rangle\langle\hat{\mathbf{e}}_2|+|\hat{\mathbf{e}}_2\rangle\langle\hat{\mathbf{e}}_1|)
\nonumber\\
&&
+\nu(|\hat{\mathbf{e}}_1\rangle\langle\hat{\mathbf{e}}_3|+|\hat{\mathbf{e}}_3\rangle\langle\hat{\mathbf{e}}_1|)
+\lambda(|\hat{\mathbf{e}}_2\rangle\langle\hat{\mathbf{e}}_3|+|\hat{\mathbf{e}}_3\rangle\langle\hat{\mathbf{e}}_2|)
\nonumber\\
&=&
s_i\mathbf{S}_i,
\end{eqnarray}
 where $s_i$ is a scalar 
\begin{eqnarray}
s_0
&=&\frac{1}{3}(\alpha+\beta+\gamma),
\\
s_1
&=&
\frac{1}{2}(\alpha-\beta),
\\
s_2
&=&
\frac{1}{2\sqrt{3}}(\alpha+\beta-2\gamma),
\\
s_3
&=&
\mu,
\\
s_4
&=&
\nu,
\\
s_5
&=&
\lambda,
\end{eqnarray}
 and
\begin{eqnarray}
\mathbf{S}_0
&=&
|\hat{\mathbf{e}}_1\rangle\langle\hat{\mathbf{e}}_1|+
|\hat{\mathbf{e}}_2\rangle\langle\hat{\mathbf{e}}_2|+
|\hat{\mathbf{e}}_3\rangle\langle\hat{\mathbf{e}}_3|=\mathbf{1},
\\
\mathbf{S}_1
&=&
|\hat{\mathbf{e}}_1\rangle\langle\hat{\mathbf{e}}_1|-
|\hat{\mathbf{e}}_2\rangle\langle\hat{\mathbf{e}}_2|,
\\
\mathbf{S}_2
&=&
\frac{1}{\sqrt{3}}\left[
|\hat{\mathbf{e}}_1\rangle\langle\hat{\mathbf{e}}_1|+
|\hat{\mathbf{e}}_2\rangle\langle\hat{\mathbf{e}}_2|-2
|\hat{\mathbf{e}}_3\rangle\langle\hat{\mathbf{e}}_3|
\right],
\\
\mathbf{S}_3
&=&
|\hat{\mathbf{e}}_1\rangle\langle\hat{\mathbf{e}}_2|+
|\hat{\mathbf{e}}_2\rangle\langle\hat{\mathbf{e}}_1|.
\\
\mathbf{S}_4
&=&
|\hat{\mathbf{e}}_1\rangle\langle\hat{\mathbf{e}}_3|+
|\hat{\mathbf{e}}_3\rangle\langle\hat{\mathbf{e}}_1|.
\\
\mathbf{S}_5
&=&
|\hat{\mathbf{e}}_2\rangle\langle\hat{\mathbf{e}}_3|+
|\hat{\mathbf{e}}_3\rangle\langle\hat{\mathbf{e}}_2|.
\end{eqnarray}
 Here, we have employed the convention for the Gell-Mann matrices \cite{GellMann:1962xb} $\bblambda_n$ and
$\mathbf{S}_1=\bblambda_3$, $\mathbf{S}_2=\bblambda_8$.
Their matrix representations for $\hat{\mathbf{e}}_1$, $\hat{\mathbf{e}}_2$, and $\hat{\mathbf{e}}_3$ are
\begin{eqnarray}
\mathbbm{S}_0
&=&
\begin{pmatrix}
1&\phantom{+}0&\phantom{+}0\\
0&\phantom{+}1&\phantom{+}0\\
0&\phantom{+}0&\phantom{+}1\end{pmatrix},
\\
\mathbbm{S}_1
&=&
\begin{pmatrix}
1&\phantom{+}0&\phantom{+}0\\
0&-1&\phantom{+}0\\
0&\phantom{+}0&\phantom{+}0\end{pmatrix},
\\
\mathbbm{S}_2
&=&\frac{1}{\sqrt{3}}
\begin{pmatrix}
1&\phantom{+}0&\phantom{+}0\\
0&\phantom{+}1&\phantom{+}0\\
0&\phantom{+}0&-2\end{pmatrix},
\\
\mathbbm{S}_3
&=&
\begin{pmatrix}
0&\phantom{+}1&\phantom{+}0\\
1&\phantom{+}0&\phantom{+}0\\
0&\phantom{+}0&\phantom{+}0\end{pmatrix},
\\
\mathbbm{S}_4
&=&
\begin{pmatrix}
0&\phantom{+}0&\phantom{+}1\\
0&\phantom{+}0&\phantom{+}0\\
1&\phantom{+}0&\phantom{+}0\end{pmatrix},
\\
\mathbbm{S}_5
&=&
\begin{pmatrix}
0&\phantom{+}0&\phantom{+}0\\
0&\phantom{+}0&\phantom{+}1\\
0&\phantom{+}1&\phantom{+}0\end{pmatrix}.
\end{eqnarray}
 Note that only $\mathbbm{S}_0$ is traceful, and
that the remaining matrices are symmetric and traceless.
The convention for the normalization is
\begin{eqnarray}
\textrm{Tr}[\mathbbm{S}_0]&=&3,
\quad
\textrm{Tr}[\mathbbm{S}_i\mathbbm{S}_j]=2\delta_{ij},
\end{eqnarray}
 for $i,$ $j=1$ through 5.

\end{document}